\documentclass[useAMS,usenatbib]{mn2e}

\usepackage{epsfig}

\newcommand{\lya}{ Ly$\alpha \;$}
\def\ergcm2s{\ifmmode {\rm\,erg\,cm^{-2}\,s^{-1}}\else
                ${\rm\,ergs\,cm^{-2}\,s^{-1}}$\fi}
 
\newcommand{\tlya}{t$_{Ly\alpha}$}

\title[Predicting the Merger Fraction of  Lyman alpha Emitters from Redshift $z \sim$ 3 to $z\sim$ 7 ]{Predicting the Merger Fraction of  Lyman alpha Emitters from Redshift $z \sim$ 3 to $z\sim$ 7}
\author[Tilvi et al 2011]
{V. Tilvi$^{1}$\thanks
{E-mail:tilvi@asu.edu},
E. Scannapieco $^{1}$,
S. Malhotra$^{1}$,
 and J. E. Rhoads$^{1}$\\
$^{1}$School of Earth and Space Exploration, Arizona State University,  Tempe, AZ 85287, USA.
}
\begin{document}


\pagerange{\pageref{firstpage}--\pageref{lastpage}} \pubyear{2002}

\maketitle

\label{firstpage}

\begin{abstract}
Rapid mass assembly, likely from mergers or smooth accretion, has been predicted to
play a vital role in star-formation in high-redshift \lya\ emitters.
Here we predict the major merger, minor merger, and smooth accreting \lya\ emitter fraction
 from $z \approx3$ to $z\approx7$
using a large dark matter simulation, and a simple physical model that is successful in
reproducing many observations over this large  redshift range.
The central tenet of this model, different from many of the earlier models, is that the star-formation in
\lya\ emitters is proportional to the mass accretion rate rather than the total halo mass.
We find that at  $z\approx3$,  nearly $35\%$ of the \lya\  emitters accrete their
mass through major (3:1) mergers, and this fraction increases to about $50\%$ at $z\approx 7$.
This imply that the star-formation in a large fraction of high-redshift \lya\ emitters is driven by mergers.
While there is discrepancy between the model predictions and observed merger fractions, some of this
difference ($\sim15\%$)  can be attributed to the mass-ratio used to define a merger in the simulation.
We predict that future, deeper observations which use a 3:1 definition of major mergers will find 
$\geq 30\% $
major merger fraction of \lya\ emitters at redshifts $\geq 3$.
\end{abstract}


\section{Introduction}

While several theoretical models of \lya\ emitters have been developed 
\citep[e.g.][]{bar04,dav06,tas06,shi07,nag08,del06,kob07,kob09,day08,sam09}
we still lack a complete
understanding of how mass assembly  and star-formation occurs in these galaxies.
It is likely that mergers are the dominant mode of mass accretion resulting in star-formation
in at least some fraction ($\geq20\%$)  of \lya\ emitters  \citep[e.g.][]{pir07,bon09,tan09}.

Recently,  \citet{til09} developed a physical model of \lya\ emitters which is successful in 
explaining many observable including number density, stellar mass, star-formation rate, and clustering
properties of \lya\ emitters from $z\approx 3$ to $z\approx 7$.
This model differs fundamentally from many of the earlier models
\citep[e.g.][]{hai99,dij07,mao07,sta07,fer08}
 in that the star-formation of \lya\  emitters, and hence their
 \lya\ luminosity is proportional to the mass accretion rate rather than the total halo mass.

In this paper, we combine the above model with a large dark matter cosmological simulation to 
predict the major merger, minor merger, and smooth accreting \lya\ emitter fraction
from $z\approx3$ to $z\approx 7$.
We also carefully asses the uncertainties in our model predictions.
We note that currently there are no theoretical prediction of merger fraction of \lya\ emitters.

On the observational front,  only recently  it  has been possible  to estimate the merger fraction of \lya\ emitters.
At $z\sim 0.3$,  \citet{cow10}
 found that $\geq30\%$   of \lya\ emitters are either irregulars or
have disturbed morphologies indicative of mergers. 
At higher redshifts, $z > 3$, the observed merger fraction varies from $\approx 20\%$ to $\approx 45 \%$ 
 \citep[]{pir07,bon09,tan09}.
 
Recently, \citet{coo10}
 found that all Lyman-break galaxies (LBGs)  that have  close companions 
($\leq 15 h^{-1}$ Mpc) exhibit  strong \lya\ emission lines. In the spectroscopically confirmed close 
pairs, the spectra/imaging show double \lya\ emission peak/double morphology confirming that these 
LBGs at $z\approx3$ are indeed mergers.
\citet{coo10}
 also studied a sample of spectroscopically confirmed LBGs at $z\approx3$ from
\citet{sha06},
and found that about 33\% of LBGs with strong \lya\ emission have double 
\lya\ emission peaks, strengthening the above conclusion.


The outline of this paper is as follows. In \S2 we briefly describe the physical model of \lya\ emitters 
\citep{til09},
 our  simulations,  and how we construct the merger trees.
In  \S3,  we estimate the merger fraction of \lya\ emitters, discuss the effect of  mass resolution in the 
simulation on our results, and compare our model predictions with the observations.
In \S4 we quantify the  uncertainties associated with the predictions, 
and we summarize our results  in \S5.

\section{Methods} 
\subsection{Modeling  \lya\ Emitters}

Our method is based on carrying out dark matter (DM) only simulations, and then  populating the halos
 from these simulations with
\lya\ emitters according to a simple physical model presented in \citet{til09}.
This model uses a  single parameter  to successfully reproducing 
many observed properties of \lya\ emitters, including their luminosity 
functions, stellar masses, stellar ages, star formation rates, and clustering from $z=3.1$ to $z=6.6$.
The central tenet of this model is that the \lya\ luminosity is proportional to the star formation rate $i.e.$
\begin{equation}
\mathrm { L_{Ly\alpha}= 1 \times 10^{42} \times \frac{SFR}{M_{\odot} { yr^{-1}}} \; \; { erg} \; { s}^{-1},}
\end{equation}
which in turn is proportional to the mass accretion rate $i.e.$  SFR =f$_\star \times \dot{M_b}.$
Here, f$_\star $ and $\dot{M_b}$ are the star formation efficiency (i.e. converting baryonic mass
to stars), and the baryonic mass accretion rate.
As we use DM only simulations, the
baryonic mass accretion rate is obtained by converting DM  mass accretion assuming
universal ratio of baryonic to DM densities.
For more details about this model, we refer the reader to \citet{til09}.

\subsection{Simulation and   Halo Catalogs}

In order to implement this method, we carried out a large
N-body DM cosmological simulation (hereafter Gadget-1024), with the GADGET2 \citep{spr05} code.
We generated the initial conditions for the simulation using 
 second-order Lagrangian Perturbation Theory  \citep{cro06,tha06}.
In this simulation we use $1024^3$ DM particles, each particle with a mass 
 M$_{p}\approx  2.7\times  10^{7}$   M$_{\odot} h^{-1}$  evolved 
in a comoving volume of (102  Mpc)$^{3}$.
Using a Friends-of-Friends (FOF) halo finder \citep{dav85},
 we identify DM halos that contain 
100 or more DM particles. 
This corresponds to a minimum halo mass   M$_{halo}\approx  2.7\times  10^{9}$   M$_{\odot} h^{-1}$.

We then generate  catalogs, for redshifts from  $z=10$ to $z=3$,  which contain positions of 
halos, their  DM mass, and unique IDs of each individual particle that belongs to a given halo.
These unique particle IDs are later used to track halos between two epochs.
Throughout this chapter we assumed a flat $\Lambda$CDM cosmology with parameters 
$\Omega_{m}$=0.233, $\Omega_{\Lambda}$=0.721, $\Omega_{b}$=0.0462,
$h$=0.71, $\sigma_{8}$=0.817 where $\Omega_{m}$, $\Omega_{\Lambda}$,
$\Omega_{b}$, $h$,  and $\sigma_{8}$ correspond, respectively, to the matter density,
dark energy density, and baryonic density in units of the critical density,
the Hubble parameter in units of 100 km s$^{-1}$ Mpc$^{-1}$, and the RMS density
fluctuations on the 8 Mpc $h^{-1}$ scale, in agreement with WMAP \citep{spe07} five year
results \citep{hin09}.

\begin{figure}
\epsfxsize=10cm
\epsfbox{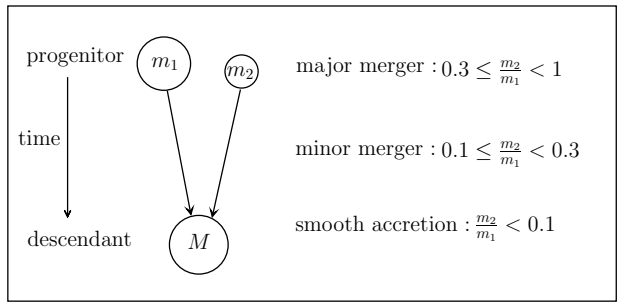}
\caption{Merger schematic showing classification of mergers based on the progenitor mass ratio.}
\end{figure}

\subsection{\lya\ Emitter Catalogs}

To construct a simulated population of \lya\ emitters, 
we first calculate the amount of DM mass that is accreted by each halo in  time
\tlya  =30 Myr \citep{til09}.
This time interval was chosen so as to match with the average stellar population
of \lya\ emitters.
This mass accretion rate is then converted to the \lya\ luminosity as described in Section 2.1. 
While, in general, we expect every halo to accrete more mass with time, we find that some halos lose mass
(negative accretion)
which is due to the limitation of halo-finding technique. 
We correct for this simulation noise by counting the number of halos that have negative mass accretion in each 
mass bin and subtracting this from corresponding counts in the positive mass bins (See \citet{til09}
for more details).
With this procedure we now have \lya\ emitter catalogs from $z=6.6$ to $z=3$.

It is possible that the  mass resolution in our simulation might affect our results.
In order to carefully assess this effect, we also used the Millenium-II simulation \citep{boy09}
which has a  larger volume, and a higher halo mass resolution.
In particular, it follows $2160^{3}$ particles with  each particle having a DM mass  
$  6.885\times  10^{6}$   M$_{\odot} h^{-1}$ in a simulation volume of (141  Mpc)$^{3}$.
Furthermore, in this simulation, each halo  is identified by an unique ID allowing us to follow
 the same procedure as applied to our Gadget-1024
simulation.
We use Millenium-II simulation only to investigate the simulation mass resolution effect since
 the time step between each output is about ten times larger,  
making it  unsuitable for our final comparisons with observations.

\subsection{Merger Tree}
As described in Section 2.1, each \lya\ emitter is assigned a luminosity based on its mass accretion rate.
This mass accretion can occur when either two or more \textit{progenitors} (halos at time step $S$-1) merge into a single
\textit{descendent} (halo at a later time step $S$), or when a single progenitor evolves into a single 
descendent (see Figure 1).

\begin{figure}
\epsfxsize=8.5cm
\epsfbox{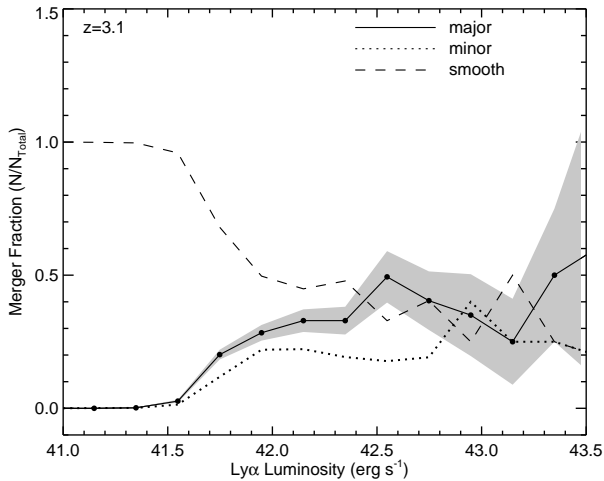}
\caption{The fraction of  \lya\ emitters that have formed from  
major mergers, minor mergers, and smooth accretion 
 at $z=3.1$.
The major and minor merger fractions are shown in solid, and dotted lines respectively, while the 
long dashed line shows the fraction of \lya\ emitters that accrete mass through smooth accretion.
The shaded area are the poisson errors on the major merger fraction. 
About 35\% of the  \lya\ emitters at $z\approx3$ accrete their mass through major mergers.
The steep decline in the major merger fraction at 
$\rm Log \; Ly\alpha < 42 \rm  \; erg \;s^{-1}$ is not real but results from limited simulation  
mass resolution (see Section 3).}
\end{figure}


We then classify each  merger into one of three categories: major merger, minor merger, or smooth 
accretion,  based on the progenitor mass ratio.
In particular,  we define each of these cases as:\\
(1) major merger: $0.3 \leq m_{2}/m_{1} \leq 1$, \\
(2) minor merger: $0.1 \leq m_{2}/m_{1}  < 0.3, $ and \\
(3) smooth accretion: $m_{2}/m_{1}  < 0.1,$\\
where  $m_{1}$ and $m_{2}$ are the masses of the most massive, and second-most 
massive progenitors associated with a single descendant (see Figure 1).
In the case where a descendent halo has a single progenitor halo, by default, this halo is assigned a smooth accretion
mode.
Using the above  criteria we  now have a merger tree, $i.e.$   each descendent halo, and hence each \lya\ emitter, 
has been associated with the
mode of mass accretion $i.e.$ major merger, minor merger or smooth accretion.

\begin{figure}
\epsfxsize=8.2cm
\epsfbox{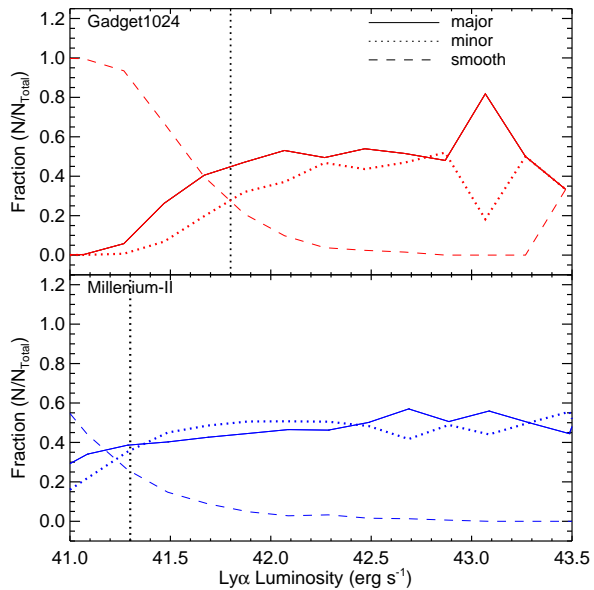}
\caption{Comparing the merger fractions between Gadget1024 and 
 the Millenium-II simulation.
The vertical dotted lines represent the luminosity range below which  our  results are not  reliable.
In the Millenium-II simulation, due to higher mass resolution, the vertical dotted line shift towards the 
lower luminosity. Also seen is the decreased noise at the brighter luminosities in the Millenium-II simulation
due to its larger simulation volume.
}
\end{figure}

\section{Model Results and Observational Comparisons}
\subsection{Merger Fraction at $z\approx 3$}

Using this method we are able to calculate the fraction of \lya\ emitters that have undergone 
major mergers, minor mergers, and smooth accretion.
Here, we remind the reader that each descendent halo hosts one \lya\ emitter.
In effect, every  descendent halo is an \lya\ emitter with its luminosity proportional to the mass accretion rate.
We define the merger fraction as the ratio of number of \lya\ emitters formed from mergers
 in a given
luminosity bin  to the total number of \lya\ emitters  in that bin. Thus,
\begin{equation}
f(L) =\frac{N_{m}(L)}{N_{T}(L) } ~~,
\end{equation}
where $N_{m}(L)$ is  the number of \lya\ emitters formed from mergers, and  $N_{T}(L)$ is the total  number of 
\lya\ emitters in that luminosity
bin.

Figure 2 shows the merger fractions of \lya\ emitters that have formed from
 major mergers, minor mergers,
and smooth accretion,   at $z=3.1$.
The major merger, and minor merger fractions are shown in solid  and dotted lines respectively, while
the long dashed line indicates the fraction of \lya\ emitters  formed from  smooth accretion.
The shaded region shows the Poisson errors on the major merger fraction.

From Figure 2 it is evident that at $z=3.1$, about  $35\%$ of \lya\ emitters are formed from major mergers.
In other words  
 the dominant mode of mass accretion, and hence star-formation  in nearly $35\%$ of 
     ${\rm Log}({\rm Ly}\alpha) > 42 \rm  \; erg \;s^{-1}$  emitters,
 occurs through major mergers, while
remaining 
 $65\%$ of \lya\ emitters accrete their mass through minor mergers and smooth accretion combined.

Note that the Poisson errors dominate the bright end of the \lya\ luminosity.
Our results might not be fully reliable below $6 \times 10^{41} \rm  erg \;s^{-1}$ where we see a steep decline in the
major merger fraction.
This effect might be partly due to  lower halo mass resolution.
To understand and quantify this effect  of lower halo mass resolution, 
 we now compare our results from figure 2 with the Millenium-II simulation
 which has about 2.6 times more volume, and nearly forty times better mass resolution.
In addition, we also investigate the effect of stellar ages ( \tlya)
 of \lya\ emitters on  our predictions.

  \begin{figure}
 \epsfxsize=8.5cm
\epsfbox{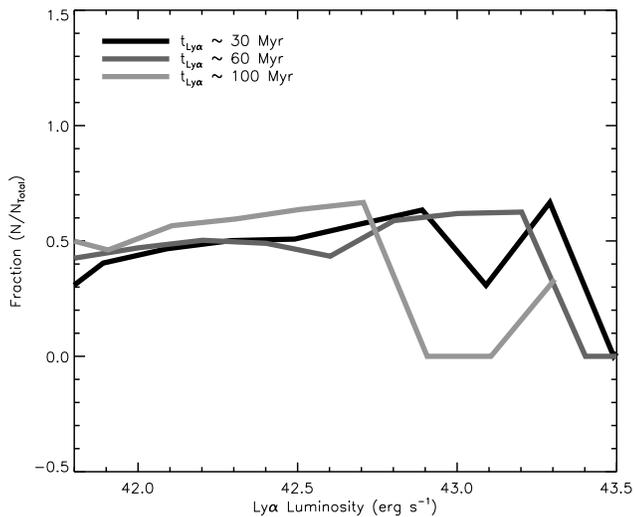}
\caption{Model prediction of major merger fraction for different stellar ages of \lya\ emitters at $z\approx7$. 
The smaller \tlya  implies younger stellar population in \lya\ emitters.}
\end{figure}

\subsection{Effect of Halo Mass Resolution and  $ t_{Ly\alpha}$ }

Using the same procedure as described in Section 2, we constructed a \lya\ emitter catalog from the 
Millenium-II simulation, at a common redshift, $z=6.8$ between the two simulations.
In Figure 3 we compare the results from the two simulations.
The top panel are the results from our Gadget-1024 simulation while the bottom panel shows the merger fraction
obtained from Millenium-II simulation.
The vertical dotted lines show the luminosity below which the  apparent decline in the major merger fraction is
caused due to lower halo mass resolution.
This luminosity range shift towards lower \lya\ luminosity in the Millennium-II simulation (lower panel) which has a
better mass resolution. Also, at the bright end of the luminosity axis, the fluctuation in the merger fraction is
lesser in the Millenium-II simulation due to the larger volume.
Thus,  better mass resolution is  needed to estimate the merger fractions reliably 
at lower luminosities, and a larger simulation volume to minimize the noise at the brighter luminosities.
Hereafter, in our analysis, we consider only those \lya\ emitters that have \lya\ luminosity 
$>6\times10^{41} \rm erg \; s^{-1}$.


We now study the effect of \tlya  on our model predicted results.
For our model \lya\ emitters the luminosity is proportional to the mass accretion rate ($\Delta M_{b}$ / \tlya)
where we assumed a fixed \tlya =30 Myr which 
corresponds to the average stellar ages of \lya\ emitters.
While this is true for average population, from observations it is inferred that this stellar age ranges from
 1 Myr to even hundred Myr  \citep[e.g.][]{fin07}.

Figure 4 shows the major merger fractions of our model \lya\ emitters at $z=6.6$ for \tlya=30, 60, and 100 Myr.
While the stellar ages have changed by a factor of 3,  we find no or little change in major merger fraction.
Thus, our model predictions are independent of the stellar ages of \lya\ emitters.

\subsection{Redshift Evolution of Merger Fraction}
In  preceding sections we have shown that the predicted merger fractions are affected by the
 mass resolution, and that there is no dependence of \tlya on the predicted merger fractions.
 We now investigate the redshift evolution of merger fraction of \lya\ emitters.

Figure 5 shows the redshift evolution of major merger fraction (left panel), minor merger fraction(middle panel),
and smooth accretion (right panel) from $z=3.1$ to $z=6.6$.
The major merger fraction (left panel),  and smooth accretion (right panel) show  
 a mild  evolution  with redshift.
 On the other hand, the minor merger fraction is nearly constant.
 At $z=3.1$ the average major merger fraction 
$f_{major}\approx 35\%$.  
 Similarly, the average fraction of minor mergers (middle panel) and smooth  accretion (right panel)
are $f_{minor}\approx 25\%$, and $f_{smooth}\approx 40\%$ respectively.
 At $z=6.6$, the major merger fraction is higher with $f_{major}\approx 50\%$.
Thus, at higher redshifts, the mass accretion and hence the star-formation in a significant fraction of the \lya\
emitters occurs through   major and minor mergers.

\begin{figure*}
\epsfxsize=18cm
\epsfbox{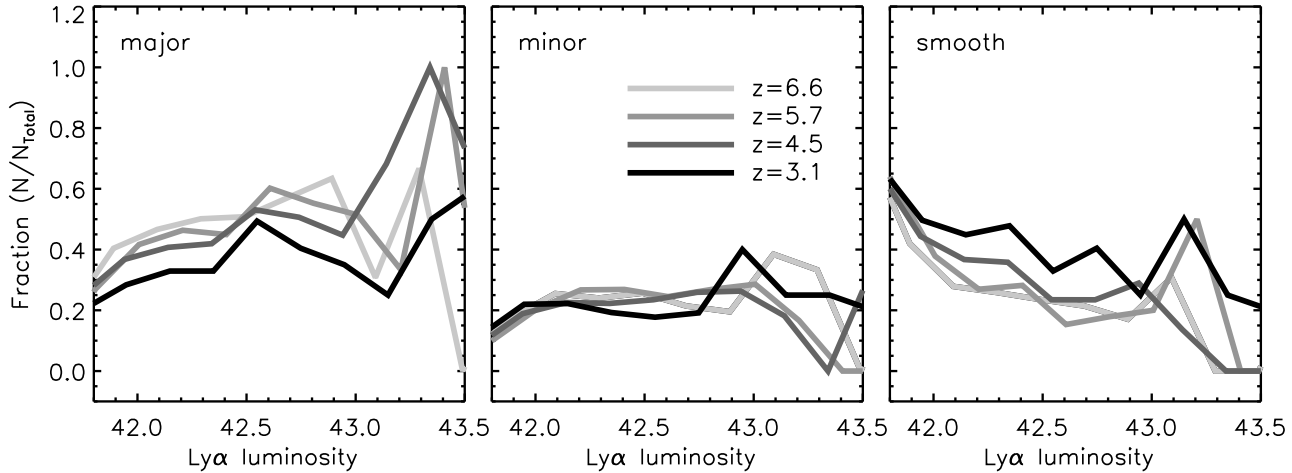}
\caption{ Redshift evolution of major merger fraction (left panel), minor merger fraction (middle), and smooth 
accretion (right panel) from $z=3.1$ to $z=6.6$.
We see a mild evolution of major merger fraction (left), and in smooth accretion (right). On the other hand the
minor merger fraction remains constant over this redshift range.}
\end{figure*}

\subsection{Comparison with the Observations}
In this section we compare only major merger fractions with the observations since it is extremely
difficult to identify minor mergers from observations especially at higher redshifts.

Only recently, has it been possible to estimate the observed merger fraction of \lya\ emitters at high-redshifts.
For example, at $z=5.7$ \citet{tan09} have studied morphological properties of \lya\ emitters in the COSMOS field using 
HST/ACS. 
Only two out of about 50 \lya\ emitters show clear signs of extended features, an indication of either
interacting or merging galaxies.
However, there is a large uncertainty in estimating  merger fraction from these observations due to their shallow 
survey depth.

In Figure 6, we compare this result with our model predictions. 
The downward arrow indicates the  upper limit on the merger fraction at $z=5.7$.
At this redshift, our model predicted major merger fraction is about $50\%$.
While comparing our model predictions with the observations we have
accounted for the observed limiting \lya\ luminosity at corresponding
redshifts.

At slightly lower redshift $z\approx 5$, \citep{pir07}
 studied morphologies of nine \lya\ emitters in HUDF.
To quantify the morphologies, they used $concentration (C)$, and $asymmetry (A)$ parameters 
\citep{con00}
and found that nearly $44\%$ have clumpy or complex structures.
Visually,  about  $33\%$ sources look morphologically disturbed or as ongoing mergers, while nearly $10\%$
of the sources can not be reliably identified as mergers.
This observed merger fraction is nearly same as  our model prediction.

\citet{bon09} studied morphological properties of about 120 $z\sim$ 3.1 \lya\ emitting galaxies in the rest-frame
ultra-violet band.
They found that at least $17\%$ of the total \lya\ emitters contain multiple components  
which might indicate 
that these are either individual star-forming regions within a single galaxy, a merged 
system or ongoing mergers.
Since it is very difficult, due to their compact sizes (e.g. Malhotra et al 2011, Bond et al 2011),  to definitely conclude whether  the multi-component \lya\ emitters 
are the remnants of mergers or if these are individual star-forming regions in a single system, we 
have shown the merger fraction by  upward and downward arrows indicating uncertainties in both directions.
Their classification is based on  counting the number of components in a fixed aperture.
 At $z=3.1$, our model predicted major merger fraction is about $35\%$, much higher than the observed
 fraction.
However, in Section 4 we show that, some of this difference between model predictions and observations can be attributed to the definition of major merger mass ratio.

At lower redshift, $z\approx 0.3$, \citet{cow10}  studied morphologies of 
\lya\ emitters in the GROTH00 and SIRTFFL00 fields.
They found that  $>30\%$ of the \lya\ emitters show signs of ongoing mergers.
Based on the above comparisons, there is some discrepancy between our model prediction and
observations. 
In the following section, we investigate the uncertainty due to the mass ratio used to define a merger.

\section{Uncertainties from Model Predictions and Observations}

\subsection{Dependence of Merger Fraction on the Progenitor Mass Ratio Definition}
We now vary the progenitor mass ratio criteria from 1:3 to 1:2, $i.e.$ major mergers are now defined 
as those for which   
  $0.5 \leq m_{2}/m_{1} \leq 1$, while for $0.1 \leq m_{2}/m_{1} < 0.5$, they are classified as minor mergers.
  All other \lya\ emitters with their progenitor halo mass ratio $m_{2}/m_{1} < 0.1$ are defined as smooth
  accreting.
 In figure 7 (see also Figure 2) we show the predicted merger fraction dependence on the progenitor mass
 ratio definition.

By changing the progenitor mass ratio from 1: 3 to 1: 2, the major (minor)  merger fraction drops (increases) by 
about $15\%$.
Thus, it is clear that the predicted merger fraction of \lya\ emitters depends on the  the progenitor mass ratio definition, and
one needs to be careful when comparing model predicted merger fractions with the observations  since the merger
defining mass ratio influences the merger fractions.

 \begin{figure}
  \epsfxsize=8.5cm
 \epsfbox{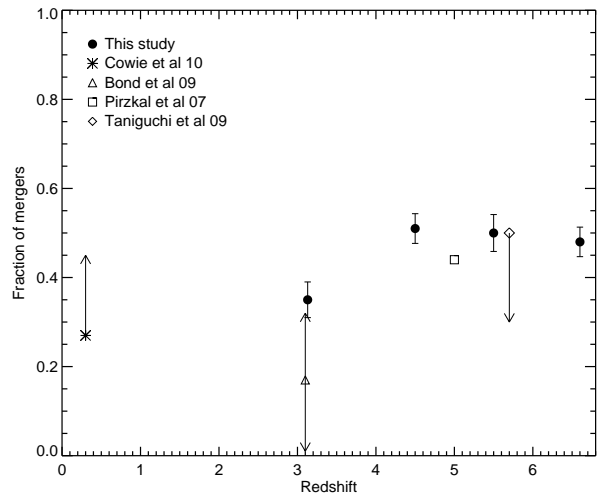}
  \caption{ Comparison of our model predicted major merger fraction with the observations.
Filled circles are our model prediction while other symbols represent observations at $z=5.7$ 
(Taniguchi et al 2009), $z\approx 5$
\citep{pir07}, $z=3.1$ \citep{bon09}, and $z\approx 0.3$ 
\citep{cow10}.  Error bars on model predictions indicate Poisson errors.}
\end{figure}



  \begin{figure}
  \epsfxsize=8.5cm
\epsfbox{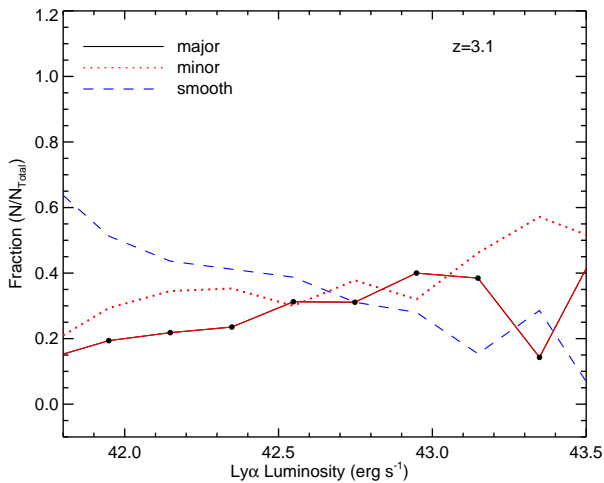}
\caption{ Progenitor mass ratio dependence on the predicted merger fraction of \lya\ emitters at $z=3.1$.
Here, the major merger is defined as $0.5 \leq m_{2}/m_{1} \leq 1$, while for the minor merger
 $0.1 \leq m_{2}/m_{1} < 0.5$. All \lya\ emitters with its progenitor mass ratio $m_{2}/m_{1} < 0.1$ are
 defined as smooth accreting.}
\end{figure}

\subsection{Merger Observability Timescale}
In addition to the above uncertainty, 
the observed merger fraction depends on the timescale, t$_{obs}$ (observability timescale)
 during which a merger can be
identified as a merger.
For example, if t$_{obs}$ is shorter than \tlya, the time during which a galaxy  is 
observed as an \lya\ emitter, then the observed merger fraction will be underestimated.
According to our model (see equation 1), merging of two galaxies results in an \lya\ emitter
with its \lya\ luminosity proportional to the mass accretion rate.
If, say  two galaxies merge on a dynamical timescale ($t_{dyn}$), then it will be observed as 
a merger for a period $t_{dyn}$=t$_{obs}$.

Using $t_{dyn}=\sqrt{3\pi/(16 G \rho(z))}$, where $\rho(z)$ is the average overdensity of the collapsing
object, we find that at  $z\approx$3,   $t_{dyn} \approx 300$ Myr, much larger than
 \tlya=30 Myr.
Thus, we expect to miss no or little merger fraction of \lya\ emitters at $z\geq3$, due to the
observability timescale.
The dynamical time, however is nearly half with  $t_{dyn} \approx  130$ Myr at $z\approx6$.

As can be seen from the above discussion that  while the predicted merger fraction is not affected by the merger observability timescale, there is some uncertainty in the predicted merger
fraction due to the defining mass ratio.
In addition, a large uncertainty is due to the shallower survey depths especially at higher-redshifts.


\section{Summary and Conclusions}
Motivated by our earlier work \citep{til09}
 based on a simple physical model of \lya\ emitters,
combined  with a large dark matter cosmological simulation, 
we have predicted the major merger, minor merger, and smooth accreting \lya\ emitter fraction
from $z\approx 3$ to $z\approx7$.
The model presented in \citet{til09}, different from many of the earlier models in that the 
star-formation rate is proportional to the mass accretion rate rather than the total halo mass, has
been  successful in reproducing
many  observed physical properties including the luminosity functions, stellar ages, star formation rates, 
stellar masses,
and clustering properties of \lya\ emitters from $z\approx3$ to $z\approx7$.

We carefully constructed the merger tree from $z\approx7$ to $z\approx3$, accounting for the uncertainties in the halo finder.
In this merger tree, each descendent halo was associated with its progenitor(s), and based on the progenitor mass
ratio, each merger was
classified as either major, minor merger or smooth accretion.
We also carefully assessed  how our predicted results might be affected by several parameters 
including   halo  mass resolution,  defining merger mass ratio, stellar ages of \lya\ emitters, and the merger
observability timescale.
We summarize our key results below:

\begin{itemize}
\item At $z\approx3$, about $35\%$ of \lya\ emitters accrete their mass through major mergers,
and this  fraction increases to about 50\% at redshift $z\approx7$. On the other hand,
at this redshift, 
the minor mergers and smooth accretion contribute equally in the remaining 50\%
of the \lya\ emitters.
Thus, at higher redshifts, mergers play an important role in mass assembly and hence the star-formation
in majority of the \lya\ emitters.
\item We also compared our model predicted major merger fraction with the observations and
found that our model  over-predicts the major merger fraction.
This discrepancy can be resolved if we take into account the uncertainties in defining the merger classification 
 mass ratio. By changing the major merger mass ratio from 3:1 to 2:1, we found that the predicted major merger
 fraction decreases by about $15\%$.
We predict that future, deeper imaging surveys and using spectroscopic methods such as the one demonstrated
in 
\citet{coo10}, should find  $\geq 30\%$ merger fraction of \lya\ emitters at high-redshifts.
\end{itemize}

This work was supported in part by the  grant HST-0808165. All simulations were performed on the $saguaro$ cluster
operated by the Fulton School of Engineering at Arizona State University.


\begin{thebibliography}{}




\bibitem[Barton et al.(2004)]{bar04} Barton, E.~J., Dav{\'e}, R., Smith, J.-D.~T., Papovich, C., Hernquist, L., 
\& Springel, V.\ 2004, ApJL, 604, L1

\bibitem[Bond et al.(2009)]{bon09} Bond, N.~A., Gawiser, E., 
Gronwall, C., Ciardullo, R., Altmann, M., 
\& Schawinski, K.\ 2009, ApJ, 705, 639 

\bibitem[Bond et al.(2011)]{bon11} Bond, N., Gawiser, E., 
Guaita, L., et al.\ 2011, arXiv:1104.2880

\bibitem[Boylan-Kolchin et al.(2009)]{boy09} Boylan-Kolchin, 
M., Springel, V., White, S.~D.~M., Jenkins, A., 
\& Lemson, G.\ 2009, MNRAS, 398, 1150 

\bibitem[Conselice et 
al.(2000)]{con00} Conselice, C.~J., Bershady, M.~A., \& Gallagher, J.~S., III 2000, AAP, 354, L21 


\bibitem[Cooke et al.(2010)]{coo10} Cooke, J., Berrier, 
J.~C., Barton, E.~J., Bullock, J.~S., 
\& Wolfe, A.~M.\ 2010, MNRAS, 403, 1020 

\bibitem[Cowie et al.(2010)]{cow10} Cowie, L.~L., Barger, 
A.~J., \& Hu, E.~M.\ 2010, ApJ, 711, 928 

\bibitem[Crocce et al.(2006)]{cro06} Crocce, M., Pueblas, S., 
\& Scoccimarro, R.\ 2006, MNRAS, 373, 369

\bibitem[Dav{\'e} et al.(2006)]{dav06} Dav{\'e}, R.,Finlator, K., \& Oppenheimer, B.~D.\ 2006, MNRAS, 370, 273 

\bibitem[Davis et al.(1985)]{dav85} Davis, M., Efstathiou, 
G., Frenk, C.~S., \& White, S.~D.~M.\ 1985, ApJ, 292, 371 


\bibitem[Dayal et al.(2008)]{day08} Dayal, P., Ferrara, A., 
\& Gallerani, S.\ 2008, MNRAS, 389, 1683 

\bibitem[Dijkstra et al (2007)]{dij07} Dijkstra, M., Wyithe, J. S. B., Haiman, Z.  2007, MNRAS, 379, 253D

\bibitem[Fernandez 
\& Komatsu(2008)]{fer08} Fernandez, E.~R., \& Komatsu, E.\ 2008, MNRAS, 384, 1363 

\bibitem[Finkelstein et al.(2007)]{fin07} Finkelstein, S.~L., Rhoads, J.~E., Malhotra, S., Pirzkal, N., 
\& Wang, J.\ 2007, ApJ, 660, 1023
 

\bibitem[Haiman 
\& Spaans(1999)]{hai99} Haiman, Z., \& Spaans, M.\ 1999, ApJ, 518, 138 

\bibitem[Hinshaw et al.(2009)]{hin09} Hinshaw, G., et al. 2009, ApJS, 180, 225 

\bibitem[Kobayashi et al.(2007)]{kob07} Kobayashi, M.~A.~R., 
Totani, T., \& Nagashima, M.\ 2007, ApJ, 670, 919

\bibitem[Kobayashi et al.(2009)]{kob09} Kobayashi, M.~A.~R., 
Totani, T., \& Nagashima, M.\ 2009, arXiv:0902.2882 


\bibitem[Le Delliou et al.(2006)]{del06} Le Delliou, M., 
Lacey, C.~G., Baugh, C.~M., \& Morris, S.~L.\ 2006, MNRAS, 365, 712

\bibitem[Malhotra et al.(2011)]{mal11} Malhotra, S., Rhoads, 
J.~E., Finkelstein, S.~L., et al.\ 2011, arXiv:1106.2816 


\bibitem[Mao et al.(2007)]{mao07} Mao, J., Lapi, A., Granato, 
G.~L., de Zotti, G., \& Danese, L.\ 2007, ApJ, 667, 655 

\bibitem[Nagamine et al.(2008)]{nag08} Nagamine, K., Ouchi, 
M., Springel, V., \& Hernquist, L.\ 2008, arXiv:0802.0228




\bibitem[Pirzkal et al.(2007)]{pir07} Pirzkal, N., Malhotra, 
S., Rhoads, J.~E., \& Xu, C.\ 2007, ApJ, 667, 49


\bibitem[Samui et al (2009)]{sam09} Samui, S. et al. 2009

\bibitem[Shapley et al.(2006)]{sha06} Shapley, A.~E., 
Steidel, C.~C., Pettini, M., Adelberger, K.~L., 
\& Erb, D.~K.\ 2006, ApJ, 651, 688 

\bibitem[Shimizu et al.(2007)]{shi07} Shimizu, I., Umemura, 
M., \& Yonehara, A.\ 2007, MNRAS, 380, L49

\bibitem[Spergel et al.(2007)]{spe07} Spergel, D.~N., et al.\ 
2007, ApJS, 170, 377	 

\bibitem[Springel(2005)]{spr05} Springel, V.\ 2005, MNRAS, 364, 1105 

\bibitem[Stark et al.(2007)]{sta07} Stark, D.~P., Loeb, A., 
\& Ellis, R.~S.\ 2007, ApJ, 668, 627

\bibitem[Taniguchi et al.(2009)]{tan09} Taniguchi, Y., et 
al.\ 2009, ApJ, 701, 915


\bibitem[Tasitsiomi(2006)]{tas06} Tasitsiomi, A.\ 2006, ApJ, 
645, 792

\bibitem[Thacker \& Couchman (2006)]{tha06} Thacker, R. J. \& Couchman, H. M. P., 2006, Int. J. High Perf. Comp. \& Net., 4, 303


\bibitem[Tilvi et al.(2009)]{til09} Tilvi, V., Malhotra, S., 
Rhoads, J.~E., Scannapieco, E., Thacker, R.~J., Iliev, I.~T., 
\& Mellema, G.\ 2009, ApJ, 704, 724




\end{thebibliography}
\end{document}